\documentclass[12pt, a4paper]{article}
\usepackage[margin=25mm]{geometry}
\usepackage[english]{babel}
\usepackage[utf8]{inputenc}
\usepackage{amsmath}
\usepackage{amsfonts}
\usepackage{graphicx}
\usepackage{latexsym}
\usepackage{graphics}
\usepackage{caption}
\usepackage{cite}
\usepackage{xcolor}
\usepackage{tikz-qtree}
\usepackage{rotating}
\providecommand{\keywords}[1]
{
  \small	
  \textbf{\textit{Keywords---}} #1
}

\title{\textbf{Stratified Mobility, Segregated Boundary, and Socioeconomic Mixing in New York}}
\author{Rafiazka Millanida Hilman  \\
        \small Department of Network and Data Science, Central European University, 1100 Vienna, Austria  \\
        \small Department of Urbanism, Delft University of Technology, 2628 CD Delft, The Netherlands  \\
 \\
}
\date{} % Comment this line to show today's date

\begin{document}
\maketitle

\begin{abstract}
Mobility cross spatial units represents the embodiment of how people manage activities between locations along temporal sequences. Spatiotemporal pattern nevertheless interacts with the socioeconomic characteristics of respected origin (push factors) and destination (pull factors) which widely discussed in spatial interaction literature. Observing this dynamics at higher spatial resolution allows us to entangle multifaceted nature of city, its complexity as a system or network, and the way it shapes movement of people. This study explore the extent interconnected elements of urban system or urban networks, in parallel with the appearance of external shock namely COVID outbreak, may affect estimation of mobility flows. To improve predictive power, gravity model is extended to urban system model by augmenting the complexities of urban network based on micro-analytical approach (intra-city networks). Our findings reveals better performance of a more complex urban system model as to compared with gravity model. Here, we leverage stratification in mobility by specifying mobility flows with respect to income status of respected areas. The occurrence of COVID outbreak followed by lockdown measure increases intra-class mobility, indicating the coupling between socioeconomic distance and geographical distance. Flows between areas with similar economic ranges are more predictable than the one of different level. Furthermore, the presence of pull factors is more affluent than push factors in determining mobility regardless the severity of external shock. 
\end{abstract} \hspace{10pt}

%TC:ignore

\section{Introduction}
\label{into}
\setlength{\parindent}{2em}
\setlength{\parskip}{1em}

Cities represent a unified boundary for various activities to take place. The way people organise their life through commuting between home, workplace, and other points of interest (POIs) \cite{noulas2012tale, schneider2013unravelling} which later followed by building social network \cite{helsley2014social, herrera2015anatomy} along temporal sequences shapes the dynamics. Common questions may arise frequently regarding how to reach a place at shortest distance and time among available combination of transportation mode or even what kind of interaction emerges from typical visiting patterns. Despite the important notion of human mobility aspect as seen in the case of migration and urbanisation \cite{andersen2011end, prieto2018gravity, sirbu2021human}, traffic and transportation \cite{nagy2018survey, ebrahimpour2019comparison}, and epidemic spreading \cite{gross2020epidemic, hazarie2021interplay},
the distribution of locations across spatial areas also dictates the urban ecosystem. Agglomeration factors used in production namely capital, infrastructure, and information drives the economic existence of cities, resulting in spatial heterogeneity of productivity \cite{fang2017urban, cottineau2019defining}. Further inquiries try to refine the empirical relationships in investigating whether higher availability and accessibility of amenities (e.g.: schools, shopping blocks, and stations), including the size of central business district and industrial cluster, guarantees the well-being of whom living there or the city itself at large \cite{pappalardo2016analytical, voukelatou2021measuring}. On top of that, emerging literature on street network highlights greater support offered by the physical construction of street and junction to facilitate urban process \cite{marshall2018street, boeing2019urban}. Its structural topology doesn't only stands as a container of pathway from one location to the rest but rather reflects the synthesis of socioeconomic, cultural, and political reflection \cite{wang2021life}. Taking a broader view, one may ask how the street network affects urban circulation systems \cite{li2018improved}, urban functional structure \cite{tomko2008experiential} and urban attractiveness \cite{wen2017understanding}.

Multifaceted nature of city builds on complexity, an object of study referenced as urban networks \cite{pflieger2010introduction, neal2012connected} or urban system \cite{albeverio2007dynamics, mcphearson2016advancing}. The growing scholarly works dedicated to pursue this line of research expands to different directions with diverse focal points, making a consolidated definition and workhouse for urban networks hard to delineate. An attempt to formulate the spectrum of urban network employs heuristic identification spanning over three layers \cite{neal2012connected}. Firstly, micro-analytical approach specifies intra-city networks, for example the impact of street network on pedestrian flows, neighbourhood walkability, and potential interaction \cite{bielik2018measuring}. Meso-level analysis comes across on the second layer by denoting regional connectivity in polycentric coalesce such as urbanisation and sub-urbanisation \cite{liu2016measuring}. At aggregate overview, macro observation captures inter-city networks, among others in the case of global port cities centrality \cite{berli2018sea}. 

Inadequacy in portraying interconnected elements in urban network or urban system could be an impediment in understanding mobility flow in city. Flow generation process aims to estimate the volume of movements between locations given the limited demographic and geometric data of respected spatial units \cite{barbosa2018human, montanari2012cross}. Gravity model is considered as the backbone in mobility flows in multiple backgrounds such as human migration \cite{anderson2011gravity} and international trade \cite{benedictis2011gravity}. It is postulated based on Newton's law of gravitation where  increase in the number of people moving between a pair of areas is in line with the population size of those areas but as distance taking them apart, decreasing flow is expected. 

Given the lack of formulation on the connectivity between mobility and the spatial context around it, modelling urban mobility is problematic and the and the question of which aspect at which scale might strongly affect the flows remains unclear.  In this research, we take a step forward to improve predictive power of gravity model within spatial interaction research by augmenting the complexities of urban network based on micro-analytical approach (intra-city networks). We construct a framework that weights urban function (e.g.: human mobility and temporal traverse) and urban form (e.g.: street topology and building footprints) in parallel because individual movement between two places bridges the interaction between them that further magnifies the spatial dynamics. With the increasing availability of heterogeneous source of data, a new chapter in modelling urban mobility is decoupled by associated dimensions: socioeconomic stratification (e.g.: income status), transportation network (e.g: subway, bus, metro, and train service), and external shock (e.g.: COVID outbreak).

It is in our interest to establish the line between the aforementioned dimensions and mobility flows as well as to measure an extent those contribute to the better prediction of spatial interaction via mobility flows in urban space. We specify three questions for this purpose: (1) How socioeconomic status and COVID epidemic spreading stratifies mobility flows across spatial units? (2) Could a better fit for mobility flow prediction be generated after taking socioeconomic stratification, transportation network, and spatial morphology into account? We take New York City as a case study to capture the changing dynamics of mobility flows in the presence of associated dimensions.

\section{Data description}
\label{data}
\setlength{\parindent}{2em}
\setlength{\parskip}{1em}

Our choice of New York City as a case study is motivated by the fact that data highlighting urban aspects are completely available at fairly granular scale (Census Tract). It serves out purpose in stepping out beyond sole mobility data by synthesising heterogeneous source of data such as socioeconomic survey, street map, transportation transit feed, and building footprints. In the following subsection we present heterogeneous source of data collections comprising mobility, socioeconomic, morphology, and transportation data.

\emph{Mobility data:} Origin-Destination (OD) Matrix is constructed based on visit trajectories of anonymous mobile phone users at large scale provided by SafeGraph \cite{kang2020multiscale}. Aggregation is conducted at census tract level with monthly temporal unit. We compare three periods that overlap with policy response on COVID outbreak. The baseline period is April 2019, representing the normal mobility level. As stay-at-home order was extended throughout April 2020 and followed by daily subway closures from 1 AM to 5 AM, this period is used as lockdown subset. A year later, mobility restriction was uplifted to a greater extent due to the arrival of vaccination program. Therefore, we also compare individual movement across census tract in the course of April 2021 to refine the mobility dynamics affected by different policy stringency.

\emph{Socioeconomic data:} Median household income is preferred as a basis for income status computation. It is recorded in American Community Survey (ACS) 5-Year Estimate Section DP03 on Selected Economic Characteristics 2019 \cite{census2019}. We create 10 classes for income status by binning the continuous numeric data of median household income into discrete quantile. This quantile-based discretisation approach results in uniform number of census tract in each bin/income class. Income status of census tract ranges from the poorest (1) to the richest (10).

\emph{Morphology data:} The first layer representing morphological structure of urban areas is street network. Conceptualised as a graph, it consists of node of which delineate intersections and dead-ends and edge as a physical substitute for street segments \cite{barthelemy2008modeling}. Raw street network data are collected from OpenStreetMap \cite{openstreetmap} and for every census tract boundary, more than 30 metrics and topological measures are calculated \cite{boeing2019street}. To get better connectivity context in street network, additional measures indicating the degree of connectivity are considered \cite{anciaes2017social} such as gamma index and alpha index. Features selection in the form of Principal Component Analysis (PCA) is performed to selectively eliminate variables with low relevance. The most informative features with ability in explaining more variances are number of street segment and intersection. On top of street network layer, building configuration layer is constructed. Building footprint data is retrieved from New York Open Data \cite{nyc_open_data}, capturing the full perimeter outline that belongs to each building as viewed from aerial imagery. Given the geographic location of each building, we identify its census tract and take the average value of roof height above ground elevation at census tract level.

\emph{Transportation data:} General Transit Feed Specification (GTFS) becomes the integrated reference for public transportation data. In its standardised format, transport operator can push the real-time transit data into static components (e.g.: schedules, fares, and stop locations) and real-time component (e.g.: trip updates and vehicle positions). GTFS is accessible from TransitFeeds as it is now an integral part of OpenMobilityData \cite{openmobilitydata}. We acquire four modals: subway (NYC Subway), metro (Metro-North Railroad), train (Long Island Rail Road), and bus (MTA Bus Company).

\section{Results}
\label{sec:res}
\setlength{\parindent}{2em}
\setlength{\parskip}{1em}

Our study is motivated by the need in modelling interaction structure between spatial units represented by census tract. The dynamics within such structure is embedded in the individual flows at temporal sequence and most likely intertwined by socioeconomic stratification and connectivity. To incorporate stratification in observing flows, we propose a framework that combines two-stages spatial interaction: between a pair of census tracts and between a pair of income classes where the census tracts belong. The mobility network is constructed as a symmetric directed network with self loops and weighted by number of people moving from one census tract to another. Formal formulation of an ordered pair $G = (N, E)$ where census tract $n$ is a node in set $N$ and the set of edges $e_{i,j} \in E$ consists of symmetric links between subset node origin census tracts $i$ and destination census tracts $j$. The weights of the network is defined as $w_{i,j}$ for $i,j \in [1, N]$. Stratification is induced by assigning a set of income classes $c_i=o \in C_I$ to each origin census tract and $c_j=d \in C_J$ to each destination census tract.

\subsection{Mobility stratification matrix}
\label{stratificationmatrix}

\begin{figure*}[ht!]
    \centering
        \includegraphics[width=0.5\linewidth]{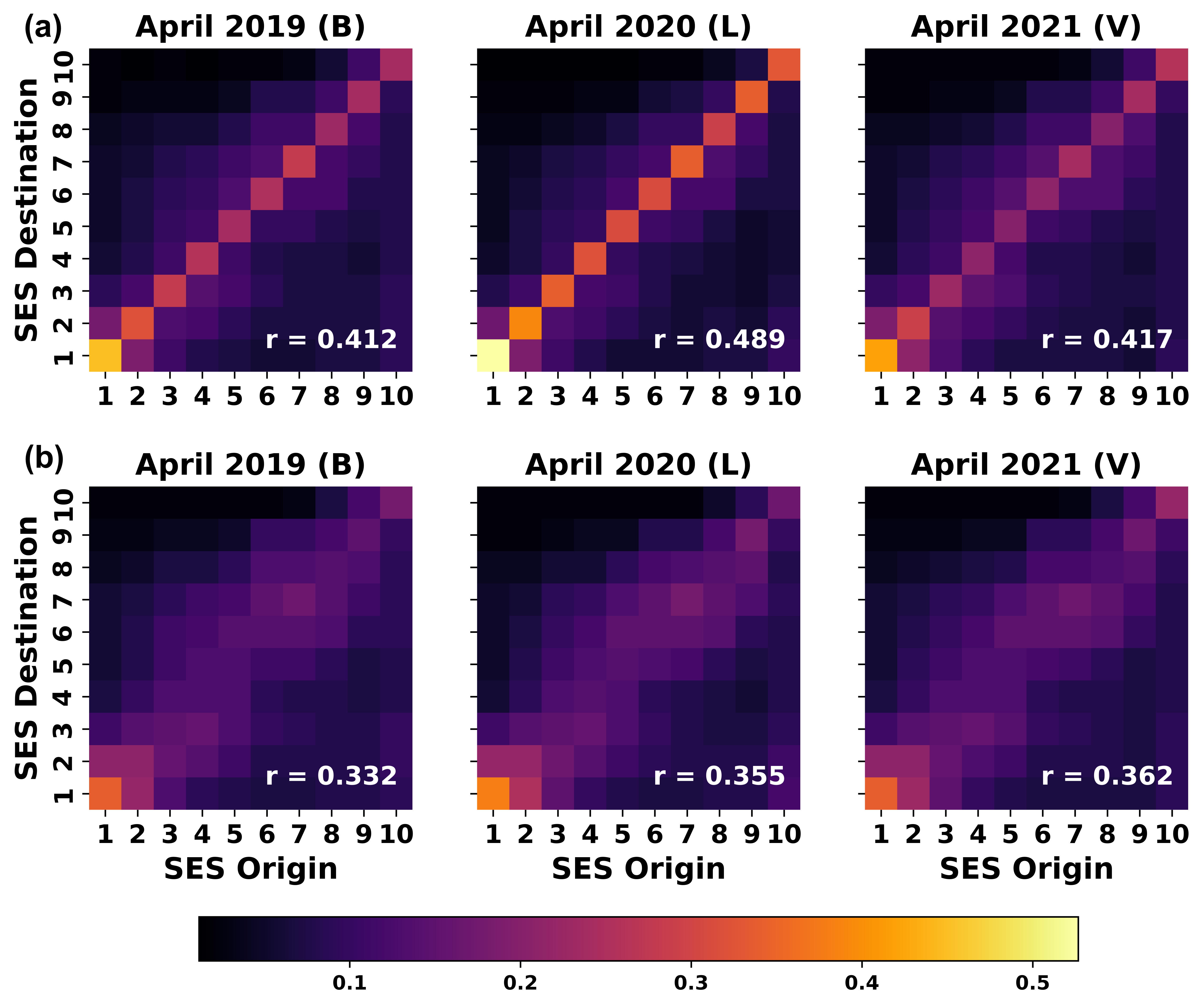}
        \includegraphics[width=0.4\linewidth]{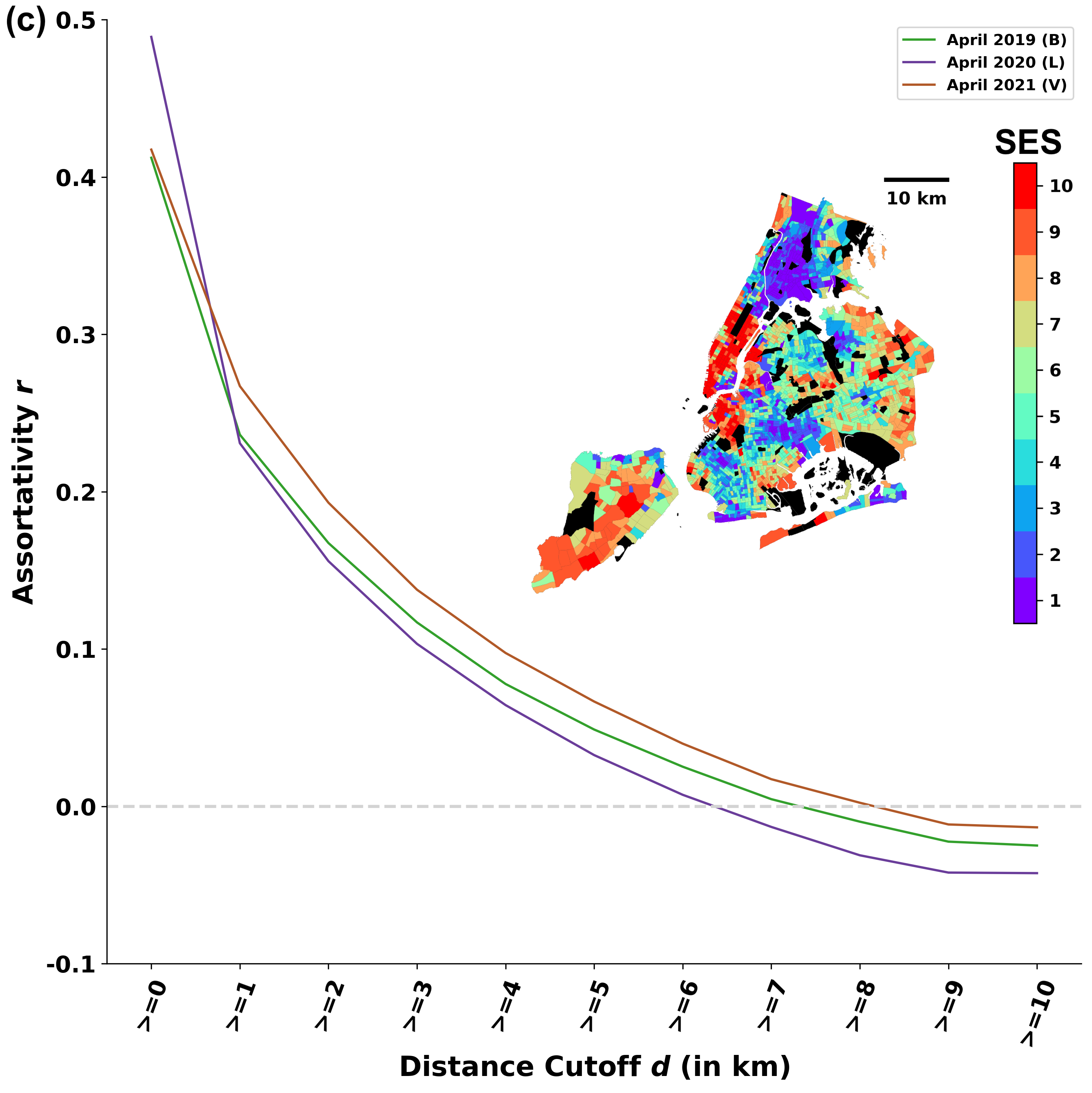}
        \caption{\textbf{Mobility stratification matrix $M_{o,d}$}. Income stratification in mobility is observable by constructing a matrix containing visit probabilities from a pair of income class between origin and destination census tract. Colour gradient indicates the magnitude where lighter colour emphasises the higher level of intensity. Fig.~\ref{fig:1}a comprises all visits, including mobility within census tract. The occurrence of COVID outbreak followed by lockdown measure increases intra-class mobility as correlation coefficient $r$ in April 2020 (0.489) is higher than baseline period in April 2019 (0.412) and vaccination period in April 2021 (0.417). Interestingly, after removing mobility within census tract as seen in Fig.~\ref{fig:1}b, similar pattern persists even with slightly higher assortativity $r$ during vaccination period. It implies that residual force of socioeconomic homophily remains in place even in the following period when mobility becomes less restricted. SES distance is intertwined with geographic distance as seen in Fig.~\ref{fig:1}c}.
        \label{fig:1}       % Give a unique label
\end{figure*}

Individual flows across census tracts can be characterised by socioeconomic inclination that simultaneously drives stratification in mobility. A \emph{mobility stratification matrix} counts for probability that a trip is made by individual departed from a census tract $i\in I$ from a given income class $c_i=o\in C_I$ to destination census tract $j\in J$ located in income class $c_j=d\in C_J$. For each pair of income class/socioeconomic status (SES) namely SES origin and SES destination, matrix element is formalised as:

\begin{equation}
M_{o,d}= \frac{\sum_{I,c_i=o}\sum_{J,c_j=d} w_{i,j}}{\sum_{d\in C_J}\sum_{I,c_i=o}\sum_{J,c_j=d} w_{i,j}}
\end{equation}

where flows from class $o$ to class $d$ are fitted into numerator and normalised by column as seen in denominator such that distribution of flow probability for each origin census tract class $o\in C_I$ could be measured. It becomes the backbone for computation in Fig.~\ref{fig:1}. Given two scopes of mobility respectively inter-census tract and intra-census tract mobility, we separately construct two matrices: all flows (Fig.~\ref{fig:1}a) and inter-census tracts flows (Fig.~\ref{fig:1}b). We take the later as refined comparison after controlling for socioeconomic distance and geographic distance by taking out intra-census tract mobility (self loops). This procedure is intended as robustness check in order to reveal the potential bias arising from short distance flows and the degree to which deviation from general mobility pattern might occur. 

To incorporate the impact of temporal dynamics driven by COVID outbreak and policy responses that come after, three sequential periods are observed independently for each mobility scope. We present findings based on mobility pattern in April 2019 as baseline (B) to be compared with period during the implementation of lockdown (L) as of April 2020 and vaccination (V) in April 2021.     

Visual representation of mobility stratification matrix exhibits prominent diagonal elements as the colours are noticeably brighter than the upper or lower diagonal. To quantify the strength of diagonal elements, we use Pearson correlation coefficient of matrix elements as an index for \emph{assortativity $r$}. It is in line with previous studies \cite{newman2003mixing, dong2020segregated, bokanyi2021universal} that specifies 

\begin{equation}
\ r = \frac{\sum_{o,d} o d M_{o,d} - \sum_{o,d}o M_{o,d}\sum_{o,d} d M_{o,d}}{\sqrt{\sum_{o,d} o^2 M_{o,d} - \left(\sum_{o,d} o M_{o,d}\right)^2}{\sqrt{\sum_{o,d} d^2 N_{o,d} - \left(\sum_{o,d} d M_{o,d}\right)^2}}}.
\end{equation}

The value of this index ranges from $-1$ (perfect disassortativity) to $1$ (perfect assortativity). If perfect disassortativity takes place, the mobility flows between origin and destination census tract will tend to be further away from similar income ranges. Consequently, more heterogeneities in spatial interaction are expected due to wider movement beyond what socioeconomic ties dictate. On the other hand, if the notion of perfect assortativity holds, the mobility flows will rather be concentrated along matrix diagonal. Furthermore, high individual mobility between two census tracts with similarities in socioeconomic level triggers more homogeneous spatial interaction. In case $r = 0$, no correlation between income class of origin and destination census tract is detected, suggesting no typical socioeconomic preferences in mobility flows.

Fig.~\ref{fig:1}a shows the persistent dominance of flows among census tract within the same income class/SES regardless the time period and policy intervention on mobility restriction. Colour gradient of the diagonal elements are visibly lighter than the rest, implying higher proportion of probability visit. Assortativity index confirms this particular circumstance as $r$ takes positive value at considerable magnitude. During period with normal mobility level (April 2019), the baseline assortativity $r = 0.412$ is already quite high, especially between census tracts at the lowest 10 percentile where at least 40\% of flows happens within the same income class. Increasing mobility isolation up to 16.69\% is found during lockdown in April 2020 with $r = 0.489$. Higher mobility intensity between census tracts in the lowest income class reaches further degree around 50\%. On this point, the extension of stay-at-home order and school closures by Governor Cuomo that was made effective on April 6, 2020 brings another boundary that enforces mobility stratification to a greater extent. The coming vaccination for three categories in April 2021 namely people with underlying conditions, age 30 and up, as well as age 16 and up makes bolder steps for reopening possible. Pupils could return to school to resume in-person learning, restaurant reopening was allowed at 25\% capacity,  so did theatres were permitted to open the door. Lower degree of mobility restriction brings back assortativity to nearly similar level to baseline period. 

To check the robustness of assortativity pattern, we compute mobility stratification matrix only for inter-census tract flows. After eliminating self loops from mobility network, such stratification pattern becomes less evident as the overall $r$ in Fig.~\ref{fig:1}b is lower for all period comparing to Fig.~\ref{fig:1}a. However, common pattern stands still in which $r$ reaches highest degree during lockdown (0.355). Mobility inclines to be 6.92\% more localised within similar socioeconomic range. Interestingly, as some mobility restriction dissolved during vaccination period, mobility even becomes more embedded within income class. There is residual force on socioeconomic aspect that might hinder individual from exploring more diverse areas. We notice that mobility restriction policy during COVID outbreak doesn't only interfere mobility in term of the physical geographical distance, but also scales up socioeconomic barrier known as SES distance. To expound the entanglement between the two effects, Fig.~\ref{fig:1}c plots the changing assortativity $r$ as a function of incremental cumulative distance between pairs of census tract. Foe every increase in distance cutoff $d$ by 1 km, we compute $r$ value. At $d \ge 0$, we take all mobility flows, corresponding to Fig.~\ref{fig:1}a. Meanwhile, at $d \ge 1$, only trips between census tracts located at least 1 km away are counted and such computation is repeated up to $d \ge 10$. For three different periods respectively April 2019 (green line), April 2020 (purple line), and April 2021 (brown line), $r$ decreases monotonically. Largest drops occur after the removal of trips that are less than 1 km in distance, evolve into more steady until $d \ge 9$, and later becomes flatter. Regime switch from assortativity to disassortativity can be seen from  $d \ge 6$ at the earliest, signifying the impact of lockdown in limiting mobility space. Downsliding lines are manifest at slower rate in absence of mobility restriction in the baseline period or with lower stringency in the vaccination period.

\subsection{Mobility stratification network}
\label{stratificationnetwork}
\setlength{\parindent}{2em}
\setlength{\parskip}{1em}

We find the noticeable complexity in mobility network owing to the aforementioned entanglement between socioeconomic, temporal, spatial dimension. To better understand the structure of stratification in mobility at finer scale, three types of mobility is proposed. The network edge $e_{i,j} \in E$ facilitates flows from origin census tract $i$ in income class $c_i = o \in C_i$ to destination census tract $j$ in income class $c_j = d \in C_i$. An edge belongs to downward mobility $e_{DM}$ if destination census tract is located in the less affluent region $d<o$. In the case a trip is made from lower income class to higher income class $d>o$, the edge is attributed to upward mobility $e_{UM}$. In addition, equal mobility $e_{EM}$ takes place for an edge connecting census tracts in the same income class $d=o$. To evaluate the relative importance of mobility type, the connectivity proportion 

%\begin{equation}
%p_{DM} = \frac{\sum_{I,c_i=o}\sum_{J,c_j=d} %w_{i,j} \in e_{DM}}{\sum_{i\in C_I}\sum_{j\in %C_J}\sum_{I,c_i=o}\sum_{J,c_j=d} w_{i,j}}
%\end{equation}

\begin{equation}
p_{DM} = \frac{\sum_{I,c_i=o}\sum_{J,c_j=d} w_{i,j} \in w_{DM}}{\sum_{I,c_i=o}\sum_{J,c_j=d} w_{i,j}}
\end{equation}

\begin{equation}
p_{UM} = \frac{\sum_{I,c_i=o}\sum_{J,c_j=d} w_{i,j} \in w_{UM}}{\sum_{I,c_i=o}\sum_{J,c_j=d} w_{i,j}}
\end{equation}

\begin{equation}
p_{EM} = \frac{\sum_{I,c_i=o}\sum_{J,c_j=d} w_{i,j} \in w_{EM}}{\sum_{I,c_i=o}\sum_{J,c_j=d} w_{i,j}}
\end{equation}

where for each mobility type, $p_{DM}$, $p_{UM}$, $p_{EM}$ is the fraction of flows by respected mobility type given ${e_{DM}, e_{UM}, e_{EM}} \in  E$ weighted by ${w_{DM}, w_{UM}, w_{EM}} \in  W$. The average size of flow across census tracts in every mobility type is 

\begin{equation}
E[s_{DM}] = \frac{\sum_{I,c_i=o}\sum_{J,c_j=d} w_{i,j} \in w_{DM}}{\sum_{I,c_i=o}\sum_{J,c_j=d} e_{i,j} \in e_{DM}}
\end{equation}

\begin{equation}
E[s_{UM}] = \frac{\sum_{I,c_i=o}\sum_{J,c_j=d} w_{i,j} \in w_{UM}}{\sum_{I,c_i=o}\sum_{J,c_j=d} e_{i,j} \in e_{UM}}
\end{equation}

\begin{equation}
E[s_{EM}] = \frac{\sum_{I,c_i=o}\sum_{J,c_j=d} w_{i,j} \in w_{EM}}{\sum_{I,c_i=o}\sum_{J,c_j=d} e_{i,j} \in e_{EM}}
\end{equation}

that corresponds to average edge strength by mobility type in the mobility network configuration. 

\begin{figure*}[ht!]
    \centering
        \includegraphics[width=1\linewidth]{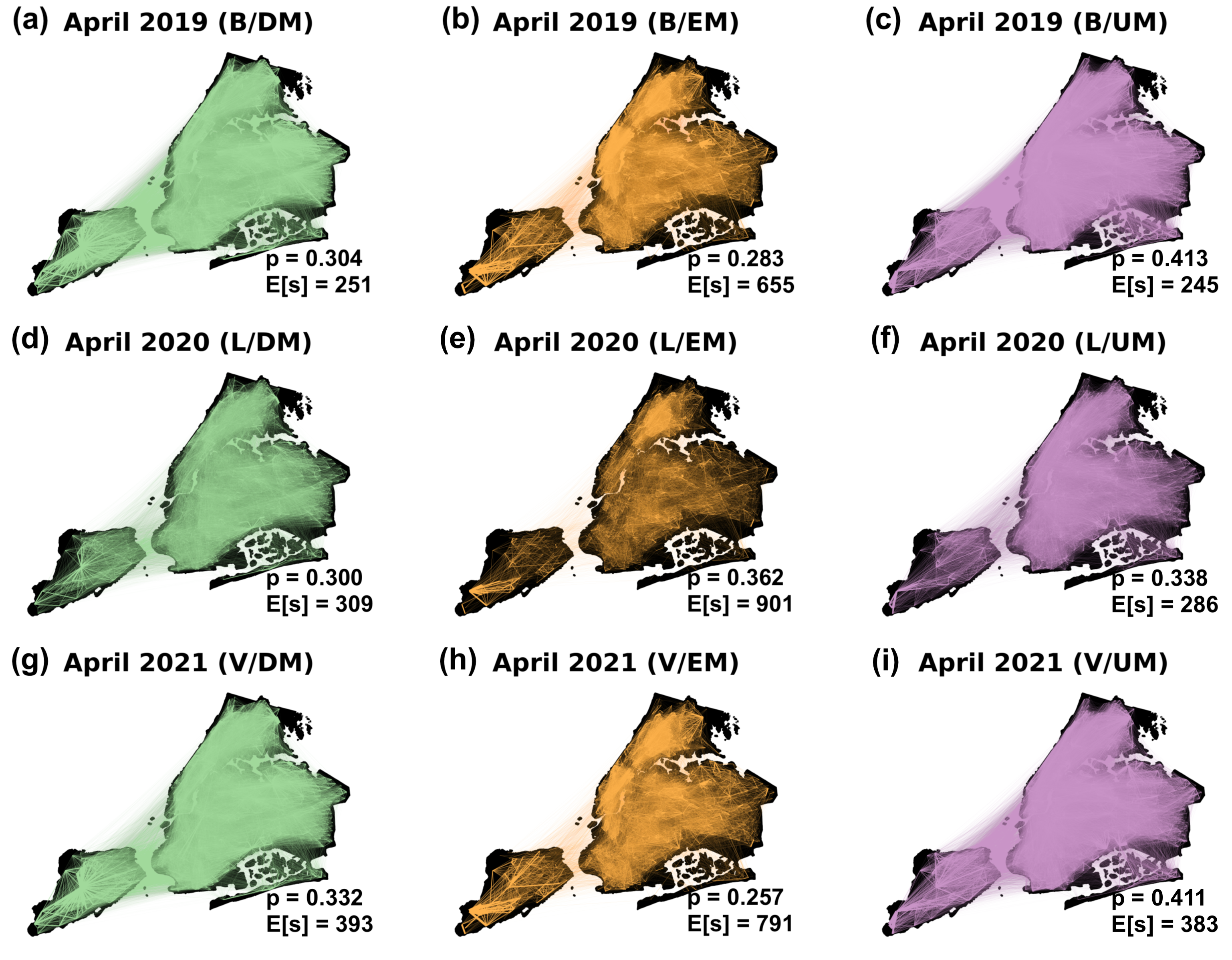}
        \caption{\textbf{Mobility stratification network}. Mobility is categorised into three distinct types. The first is downward mobility (DM/green) for which income status of destination census tract is lower than origin census tract (Fig.~\ref{fig:2}a, d, and g). In contrast, as people visit a census tract located in higher income status than the origin census tract, upward mobility (UM/purple) holds (Fig.~\ref{fig:2}c, Fig.~\ref{fig:2}f, and Fig.~\ref{fig:2}i). If the flows happen between census tract in the same income status, such mobility is denoted as equal (EM/orange) as in Fig.~\ref{fig:2}c, Fig.~\ref{fig:2}f, and Fig.~\ref{fig:2}i. Temporal disaggregation shows that upward mobility is dominant during the baseline period (April 2019), toppled down by equal mobility during lockdown (April 2020), and bounced back during vaccination (April 2021)}.     
        \label{fig:2}       % Give a unique label
\end{figure*}

Separate observation is provided in Fig.~\ref{fig:2} for every possible combination between socioeconomic and temporal dimension of mobility in order to clarify the stratification configuration. Row-wise order pinpoints the difference by mobility type (downward mobility/green; equal mobility/orange; and upward mobility/purple) within same period, while column-wise plot distinguishes temporal shift within mobility type. As for the connectivity proportion $p$, the sum over row equals to 1 and the average size of flow across census tracts for each mobility type in the designated period is denoted as $E[s]$. 

During the baseline period (April 2019), upward mobility (Fig.~\ref{fig:2}c) makes up the largest proportion as 41.30\% of trips oriented to more affluent areas. It implies the resonance of push-factor such as work places, economic opportunities, and amusement facilities that are commonly located in areas with higher income profile. The dominance of upward mobility doesn't last once lockdown is imposed in April 2020. The order flips to 36.2\% equal mobility (Fig.~\ref{fig:2}f) and followed by 33.8\% equal mobility (Fig.~\ref{fig:2}e) and 30\% downward mobility (Fig.~\ref{fig:2}d). Decreasing upper mobility is compensated by increasing equal mobility and relatively constant downward mobility. It is consistent with higher assortativity $r$ in Fig.~\ref{fig:1} as mobility is very much localised during lockdown. Configuration of mobility stratification during vaccination period in April 2021 returns to the original pattern before COVID outbreak. Upward mobility is on the top with 41.1\% share.

However, looking at the average flow ($E[s]$), mobility between areas with the same income status comes with almost 2.5 times higher intensity than any other mobility type ($E[s_{EM}] = 655$ compared to $E[s_{UM}] = 245$ and $E[s_{DM}] = 251$) as seen in Fig.~\ref{fig:2}b. It is yet more intensified during lockdown in April 2020, reaching out 37.56\% increase to $E[s_{EM}] = 901$, and levelling off to $E[s_{EM}] = 791$ in April 2021. Mobility restriction as a response to pandemics induces higher closeness between areas within similar economic ranges as a consequence of diminishing variety of mobility purpose (e.g.: minimised trips to work places and social activities). Higher homogeneity in mobility comes out as result  and different mobility type reacts to changing restriction at different rate. Therefore, it is necessary to fit this condition in modelling mobility flows.

%\subsection{Spatial morphology and spatial assortativity}
%\label{spatialmorphology}
%\setlength{\parindent}{2em}
%\setlength{\parskip}{1em}

\subsection{Mobility flow estimation}
\label{flowestimation}
\setlength{\parindent}{2em}
\setlength{\parskip}{1em}

Movement of people can't be solely perceived as physical exploration from one location to many others. It might also indirectly reflect socioeconomic ties, income inequality, and geographic constraint that encourage an individual to be present in various places. At multiple scales, aggregation over individual generates flows. Simultaneously, spatial unit takes various shapes and sizes. Aggregation over locations falls into numerous bounding boxes such as block, census tract, and borough. Ultimately, projecting mobility flows into scaled spatial unit, represented by census tract in this study, allows us to investigate the dynamics of spatial interaction. As one of the well-established frontier in the study of spatial flows, spatial interaction model offers methods in modelling relational data that are impeded by distances in geographic space. 

Retrieving its long standing theoretical development, the initial formulation of spatial interaction model, known as gravity model, is simply a translated version of Newton's law of gravitation in which particle interactions are drifted due to gravitational constant, mass of those particles, and squared distance between them \cite{roy2003spatial}. In the pioneering works, interaction is defined as number of people moving between two spatial units, mass is replaced by population size of those respected spatial units, and distance remains intact with squared effect \cite{jq1941inverse}. Determination of  parameter such as distance in the later literature shifts from rigid priory exponent -2 to be more empirical based on calibrated observation in the probabilistic framework of flows between a pair of locations \cite{huff1963probabilistic}. Calibration is usually performed under Ordinary Least Squares (OLS) regression. A technical issue arises from this part since OLS requires normality assumption regarding data distribution which is hard to fit on mobility data. As count data by nature with the possibility of zero flows and not normally distributed most of the time, we consider Generalised Linear Model (GLM) regression for Poisson distribution to address this incompatibility.  

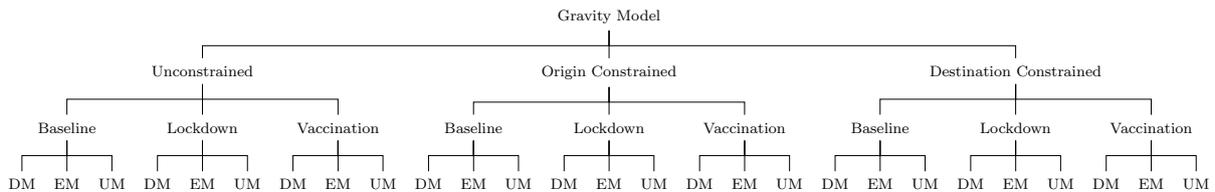
\begin{figure*}[h]
  \centering
  %\begin{turn}{90}
  \resizebox{1\textwidth}{!}{%
  \begin{tikzpicture}[font=\scriptsize]
  \tikzset{edge from parent/.style={draw,edge from parent path={(\tikzparentnode.south)-- +(0,-8pt)-| (\tikzchildnode)}}}
  \Tree [.{Gravity Model}
  [.Unconstrained  
  [.Baseline [.DM ] [.EM ] [.UM ]] 
  [.Lockdown [.DM ] [.EM ] [.UM ]]
  [.Vaccination [.DM ] [.EM ] [.UM ]]]
  [.{Origin Constrained}
  [.Baseline [.DM ] [.EM ] [.UM ]] 
  [.Lockdown [.DM ] [.EM ] [.UM ]]
  [.Vaccination [.DM ] [.EM ] [.UM ]]]
  [.{Destination Constrained}
  [.Baseline [.DM ] [.EM ] [.UM ]] 
  [.Lockdown [.DM ] [.EM ] [.UM ]]
  [.Vaccination [.DM ] [.EM ] [.UM ]]]]]
  \end{tikzpicture}}
  %\end{turn}
\caption{\textbf{Overview of gravity model configuration.} In estimating intensity of interaction between origin and destination areas represented by mobility flows, it takes into account their population and income given the restraining distance between their centroids.}
\label{fig:3a}  
\end{figure*}

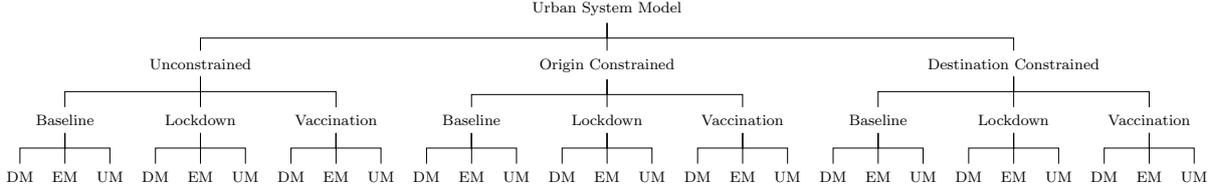
\begin{figure*}[h]
  \centering
  %\begin{turn}{90}
  \resizebox{1\textwidth}{!}{%
  \begin{tikzpicture}[font=\scriptsize]
  \tikzset{edge from parent/.style={draw,edge from parent path={(\tikzparentnode.south)-- +(0,-8pt)-| (\tikzchildnode)}}}
  \Tree [.{Urban System Model}
  [.Unconstrained
  [.Baseline [.DM ] [.EM ] [.UM ]] 
  [.Lockdown [.DM ] [.EM ] [.UM ]]
  [.Vaccination [.DM ] [.EM ] [.UM ]]]
  [.{Origin Constrained}
  [.Baseline [.DM ] [.EM ] [.UM ]] 
  [.Lockdown [.DM ] [.EM ] [.UM ]]
  [.Vaccination [.DM ] [.EM ] [.UM ]]]
  [.{Destination Constrained}
  [.Baseline [.DM ] [.EM ] [.UM ]] 
  [.Lockdown [.DM ] [.EM ] [.UM ]]
  [.Vaccination [.DM ] [.EM ] [.UM ]]]]]
  \end{tikzpicture}}
  %\end{turn}
\caption{\textbf{Overview of urban system model configuration.} On top of features used in gravity model, urban system model augment additional aspect of urban morphology (e.g.: number of intersections, street segments, and average roof height) and urban connectivity (e.g.: the availability of subway, bus, metro, train facilitating mobility) in estimating intensity of interaction between origin and destination areas represented by mobility
 flows.}
\label{fig:3b}  
\end{figure*}

The phenomena of interest in this research is mobility flows represented by given number of people moving across spatial units at census tract level and driving forces behind such mobility (e.g: economic opportunity, transportation infrastructure, and urban morphology). These factors are non-uniformly distributed across geographic spaces. In this context, mobility is highly dynamic and depends on the strength of those mobility driving forces over period of time. To improve the predictive power of spatial interaction model, at first we initialise \emph{Gravity Model} (Fig.~\ref{fig:3a}) by taking observed flows for every pair of census tract along with their population and median household income. The second step is extending the principal model to \emph{Urban System Model} (Fig.~\ref{fig:3b}) by augmenting related variables as proxy of urban morphology (e.g.: number of intersections, street segments, and average roof height) and urban connectivity (e.g.: the availability of subway, bus, metro, train facilitating mobility). This procedure reveals the contribution of each feature (basic gravity, urban morphology, and urban connectivity) in estimating empirical flows.\newline

Model parameter estimation is conducted under entropy maximising framework \cite{wilson1971family} by making use of constraints namely total inflow (number of incoming people coming) and total outflow (number of outgoing people coming) at every spatial unit. In mobility network, total inflow is in parallel with weighted in strength and total outflow is in line with weighted out strength. Decision on where to designate the constraint results in four distinct configurations: unconstrained (Section \ref{unconstrained}), origin/production constrained (Section \ref{oconstrained}), and destination/attraction constrained (Section \ref{dconstrained}). The first only considers the equivalency between observed and estimated total flows at system level (sum of all spatial units). The later enforces such condition at spatial unit level which differs in term of which set of area being controlled, the origin or destination spatial units. We prefer this approach to identify characteristics of location that play substantial role in shaping mobility with regard to push (emissiveness) and pull (attractiveness) effect.  

In addition, we expect that complexity of urban mobility at micro analytical scale (intra-city mobility) could be clarified by specifying the extent spatial context including morphology and transportation is inherent to mobility flows. We also construct separate model estimation for each mobility type (downward mobility, equal mobility, and upward mobility) in order to compare the sensitivity of socioeconomic stratification and identify which mobility type highly affected by the presence of disturbance in urban system (denoted by change in policy response to COVID outbreak). Section ~\ref{stratificationnetwork} is in line with this reasoning, therefore the following results could validate the existing finding.

\subsubsection{Unconstrained model}
\label{unconstrained}

Unconstrained model guarantees the summation of estimated mobility flows to be equal with aggregate value of observed mobility flows at system level ($T = \sum_i \sum_j T_{ij}$). This would yield multiplicative form: 

\begin{equation}
T_{i,j} = kV_i^\mu W_j^\alpha d_{i,j}^{\text{--}\beta}
\end{equation}

where $T_{i,j}$ denotes observed mobility flows between origin census tract $i$ to destination census tract $j$; $V_i$ is a origin attributes vector representing the emissivity/push factors of all origins $i \in I$ in the system; $W_j$ is a destination attributes vector representing the attractiveness/pull factors of all destinations $j \in J$ in the system; $d_{i,j}$ is a distance matrix constituting the cost that might restrain mobility flows between $i$ and $j$. Additionally, there are four model parameters to be estimated namely $k$ (proportionality constant that forces equality condition at system level aggregation); $\mu$ (scaling effect of push factor); $\alpha$ (scaling effect of pull factor); and $\beta$ (scaling effect of distance). Negative value of $\beta$ is expected because increasing distance rather discourages mobility at larger scale/wider space. 

The equation above can be rearranged in the additive form as a poisson regression:
\begin{equation}
\lambda_{i,j} = exp(k + \mu \: ln \: V_i + \alpha \: ln \: W_j - \beta \: ln \: d_{i,j})
\end{equation}

for which $\lambda_{i,j}$ is the mean of flows between $i$ and $j$ drawn from Poisson distribution ($\lambda_{i,j} = T_{i,j}$). It is logarithmically modelled by the logged independent variables in a linear combination, yielding unconstrained poisson log-linear
gravity model. Estimates of $k$, $\mu$, $\alpha$, and $\beta$ are calibrated by fitting this model with the maximum likelihood estimation method namely GLM. The built-in iteratively weighted least squares (IWSL) in GLM aims to converge parameter estimates convergence to the maximum likelihood estimates \cite{nelder1972generalized}.

\clearpage

\begin{figure*}[ht!]
    \centering
        \includegraphics[width=1\linewidth]{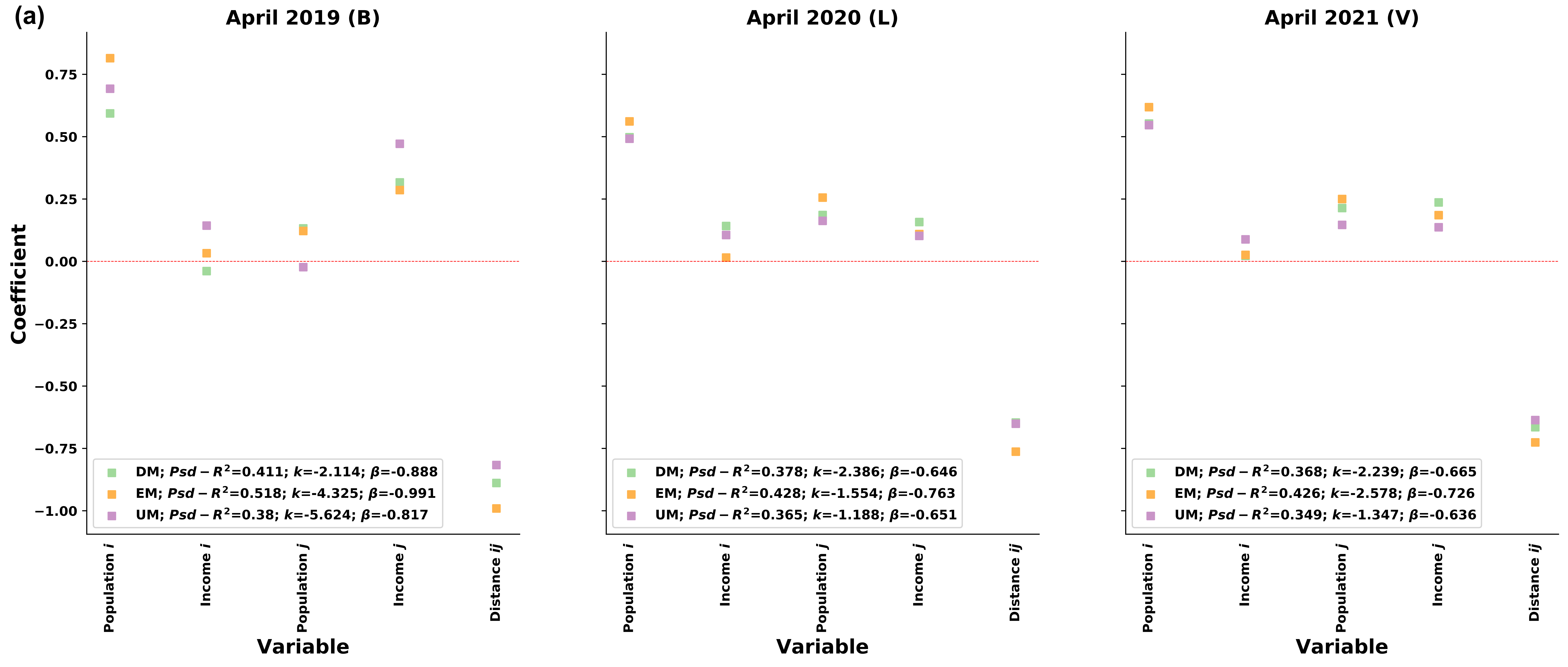}
        \includegraphics[width=1\linewidth]{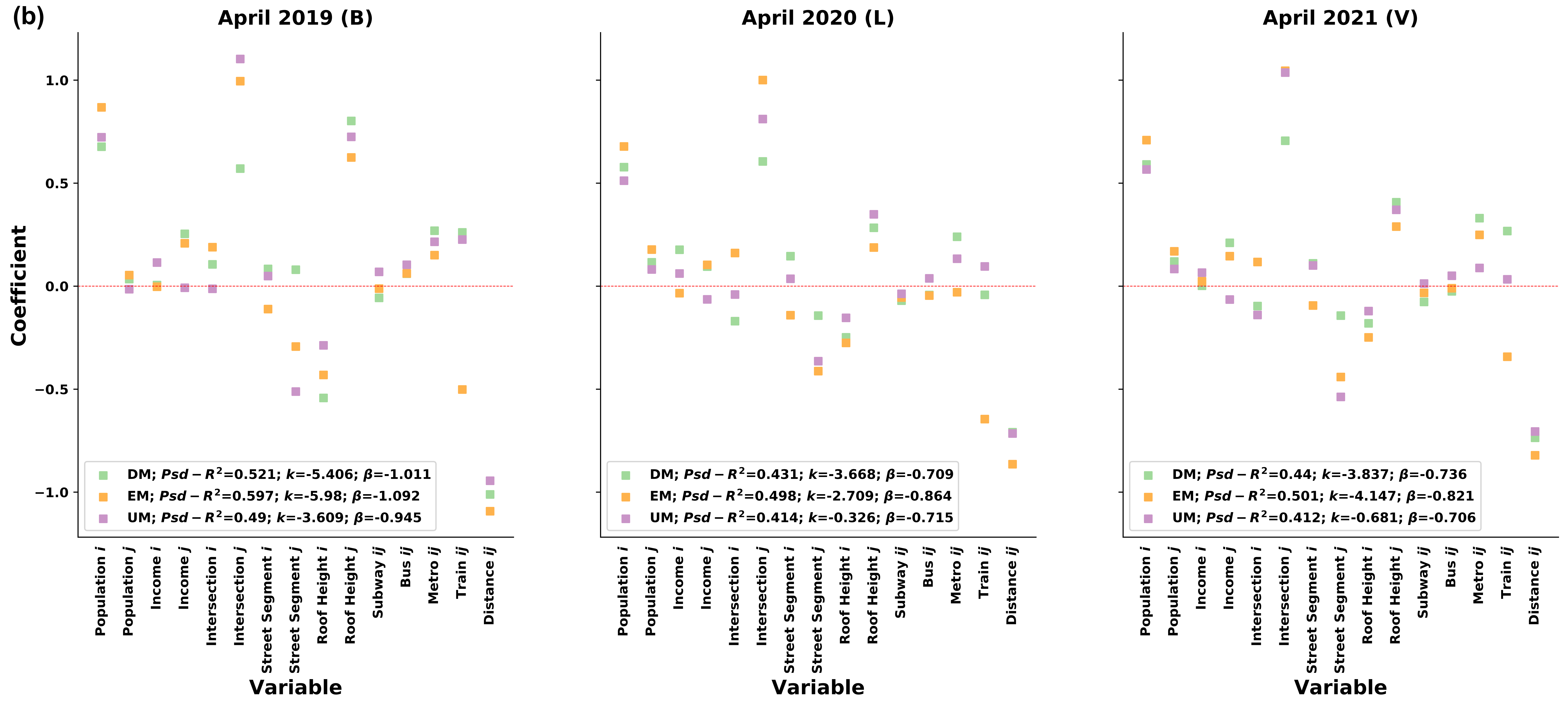}
        \caption{\textbf{Unconstrained Spatial Interaction} is formulated by assuming equal size of estimated total flows and observed total flows. Gravity Model (Fig.~\ref{fig:3}a)  could be extended into Urban System Model (Fig.~\ref{fig:3}b) for each period. The square blocks in each figure represent coefficient in the regression equation with intercept $k$ and its colour specifies mobility types (green/DM/downward mobility; orange/EM/equal mobility; and purple/UM/upward mobility). All variables are statistically significant at $\alpha$ = 5\%. Urban System Model performs better as indicated by higher pseudo $R^2$ for all mobility types. The largest predictive power improvement by 28.9\% is found in upward mobility during April 2019. In general, mobility between areas in the same income status is more predictable than others, especially before COVID outbreak (pseudo $R^2$ = 0.518 for Gravity Model; pseudo $R^2$ = 0.597 for Urban System Model). Moreover, the effect of distance (distance decay parameter $\beta$) is the strongest determinant in dictating mobility.}
        \label{fig:3}       % Give a unique label
\end{figure*}

Performance of gravity model (Fig.~\ref{fig:3}a) is compared to urban system model (Fig.~\ref{fig:3}b) under unconstrained capacity setting. In both models, all variables are also statistically significant at $\alpha= 5\%$. Distance is the strongest determinant in dictating mobility as distance decay parameter $\beta$ has the largest absolute value regardless the mobility types.  However, the implementation of lockdown results in shrinking $\beta$ value, making distance effect becomes less poignant since mobility tends to be more geographically more localised in April 2020. Downward mobility is the one largely hit as $\beta$ is dropped by 37.46\%, followed by equal mobility (29.88\%) and upward mobility (25.50\%). In urban system model, the pattern between reversed between upward mobility (32.17\%) and equal mobility (26.39\%) while downward mobility (42.60\%) still hardly hit. Furthermore, additional origin and destination attributes representing urban morphology magnifies the the pull factors, making destination area more influential in attracting flows. Among public transportation, metro is more prominent in bridging the distance due to its wide network throughout the boroughs in New York. In general, mobility between areas in the same income status is more predictable than others, especially before COVID outbreak (pseudo $R^2 = 0.518$ for Gravity Model; pseudo $R^2 = 0.597$ for Urban System Model). Therefore, prediction improvement due to feature extension to urban system model obtains highest level in baseline period (April 2019) especially for upward mobility.

\subsubsection{Origin/production constrained model}
\label{oconstrained}

Suboptimality might arises in unconstrained model due to its loose assumption on equal values between observed and empirical flows at system level instead of location level. To deal with this, one alternative to opt is constraining estimated flows to the existing outgoing flows from origin locations. It motivates the initialisation of origin/production constrained model. Moreover, constraining model estimates to relevant origin attributes could refine the analysis such as capturing the effect of higher accessibility of street in the destination area on flows coming from origin area. 

With respect to observed outflows from origin locations and the available information about them, estimated outflows are ensured to equal this capacity constraint. On the technical ground, constant $k$ at system level is longer necessary and should be replaced by balancing factor $A_i = \frac{1}{\sum_j W_j^\alpha d_{i,j}^{\text{--}\beta}}$ at location level. As a vector of values containing relevant information for mobility from each origin location, $A_i$ assures that estimated flows from every single origin is comparable to total observed flows $O_i = \sum_j T_{ij}$, not only system level aggregate. This conceptualisation generates multiplicative form: 

\begin{equation}
T_{i,j} = A_i O_i W_j^\alpha d_{i,j}^{\text{--}\beta}
\end{equation}

Holding logarithmic function that links variables to the flows $T_{i,j}$ characterised by poisson distributed mean $\lambda_{i,j}$, the equation above is transformed to:

\begin{equation}
\lambda_{i,j} = exp(\mu_i + ln \: W_j - \beta \: ln \: d_{i,j})
\end{equation}

where $\mu_i$ is a substitute to balancing factor $A_i$ serves as categorical predictor/dummy variable representing origins in the regression. 

\clearpage

\begin{figure*}[ht!]
    \centering
        \includegraphics[width=1\linewidth]{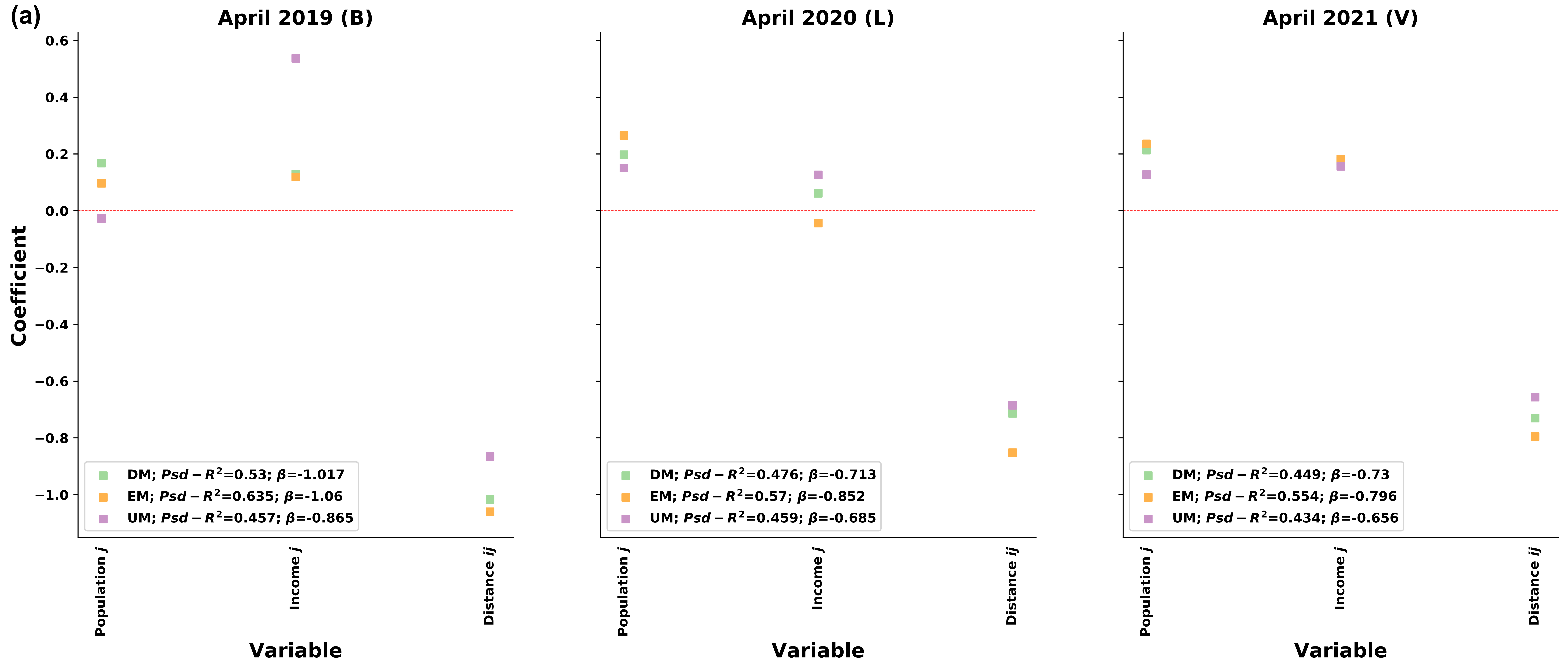}
        \includegraphics[width=1\linewidth]{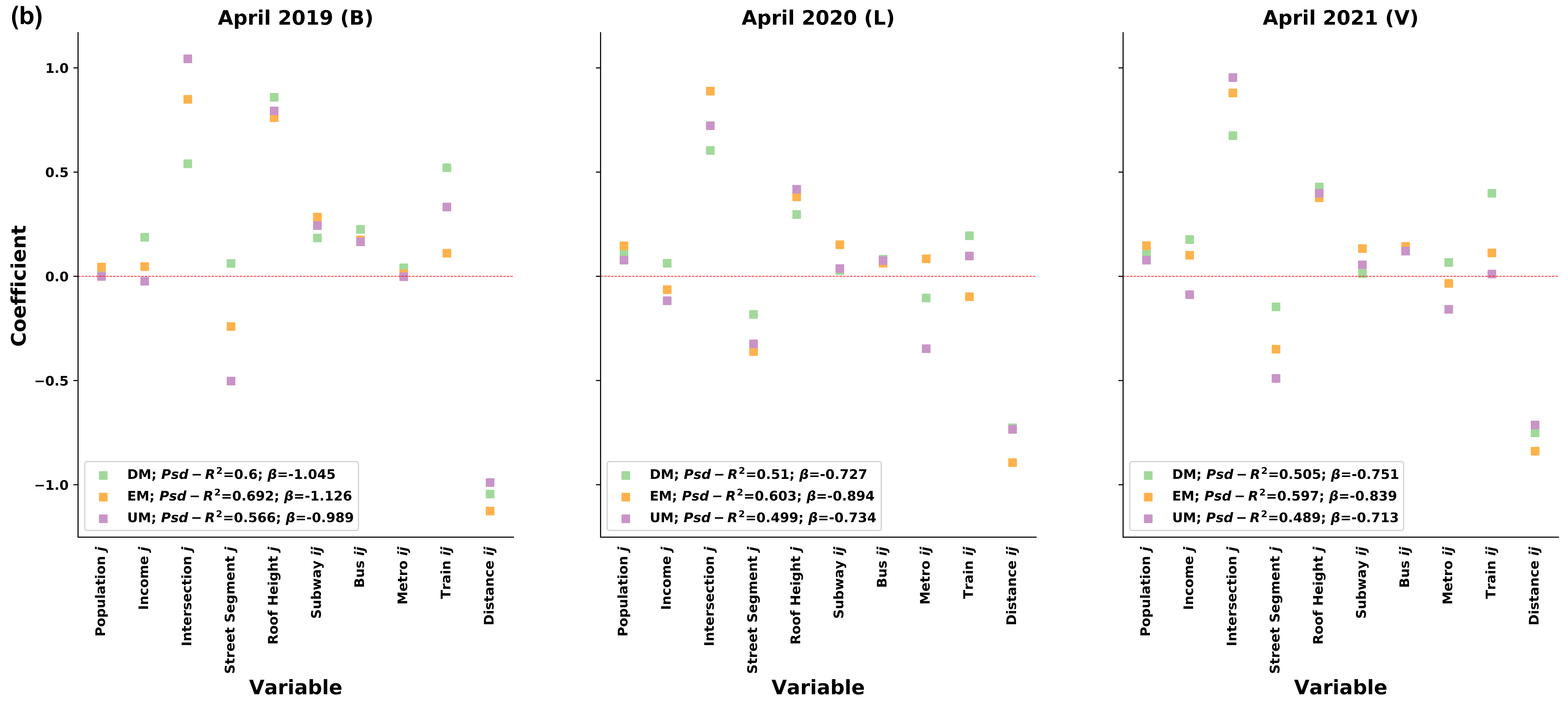}
        \caption{\textbf{Origin/Production Constrained Model} is specified by fitting origin balancing factor $A_i$ to estimated flows from each origin in such a way that its total value equals to observed total flows at origin. Extending Gravity Model (Fig.~\ref{fig:4}a) to Urban System Model (Fig.~\ref{fig:4}b) levels up pseudo $R^2$ regardless the period and mobility types with the highest increase of predictive power up to 64.3\% during April 2021.  Mobility within similar income range (EM) in the baseline period remains more predictable than any other type for both  models (pseudo $R^2$ = 0.635 for Gravity Model; pseudo $R^2$ = 0.692 for Urban System Model). All variables are statistically significant at $\alpha$ = 5\%.}
        \label{fig:4}       % Give a unique label
\end{figure*}

We compare the performance of gravity model (Fig.~\ref{fig:4}a) and urban system model (Fig.~\ref{fig:4}b) in the constrained capacity setting on origin side. In both models, all variables are also statistically significant at $\alpha= 5\%$. After controlling for origin characteristics, distance decay parameter $\beta$ becomes higher in both gravity model and urban system model, across period and for all mobility types. As being so, distance effect is  actually larger in discouraging people to explore wider areas than before. Predictability of mobility between areas with similar income range is higher and most notably in the baseline period (pseudo $R^2 = 0.635$ for Gravity Model; pseudo $R^2 = 0.692$ for Urban System Model). Interestingly, prediction improvement due to feature extension to urban system model is at its highest in vaccination period (April 2021) for upward mobility (23.85\%), highlighting the centrality of pull factor attributes.

\subsubsection{Destination/attraction constrained model}
\label{dconstrained}

On the other spectrum, constraining estimated flows to the observed incoming flows to destination locations, noted as destination/attraction constrained model, is also possible to perform. On top of that, constraining model estimates to relevant destination attributes may reveal the impact on street infrastructure in the origin area on mobility magnitude to destination area. 

In this configuration, we have by balancing factor $B_j = \frac{1}{\sum_i V_i^\mu d_{i,j}^{\text{--}\beta}}$ at location level. As a vector of values containing relevant information for mobility to each destination location, $B_j$ guarantees the equality between estimated flows to every single destination and total observed incoming flows $D_j = \sum_i T_{ij}$, not only system level aggregate. This conceptualisation generates multiplicative form: 

\begin{equation}
T_{i,j} = D_j B_j V_i^\mu d_{i,j}^{\text{--}\beta}
\end{equation}

Assuming logarithmic function that links variables to the flows $T_{i,j}$ characterised by poisson distributed mean $\lambda_{i,j}$, the equation above is equivalent to:

\begin{equation}
\lambda_{i,j} = exp(\alpha_i + ln \: V_i - \beta \: ln \: d_{i,j})
\end{equation}

where $\alpha_i$ is similar to balancing factor $B_j$ serves as categorical predictor/dummy variable representing destination in the regression. 

\clearpage

\begin{figure*}[ht!]
    \centering
        \includegraphics[width=1\linewidth]{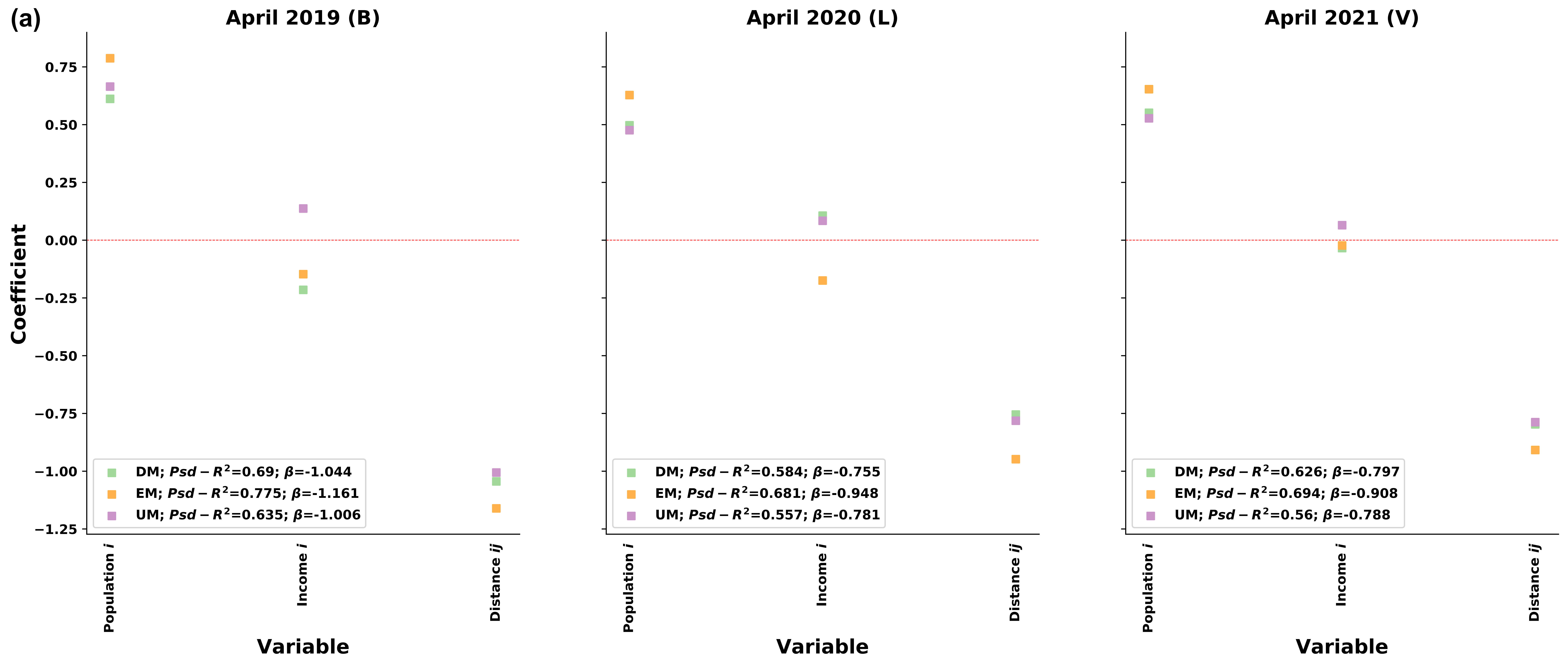}
        \includegraphics[width=1\linewidth]{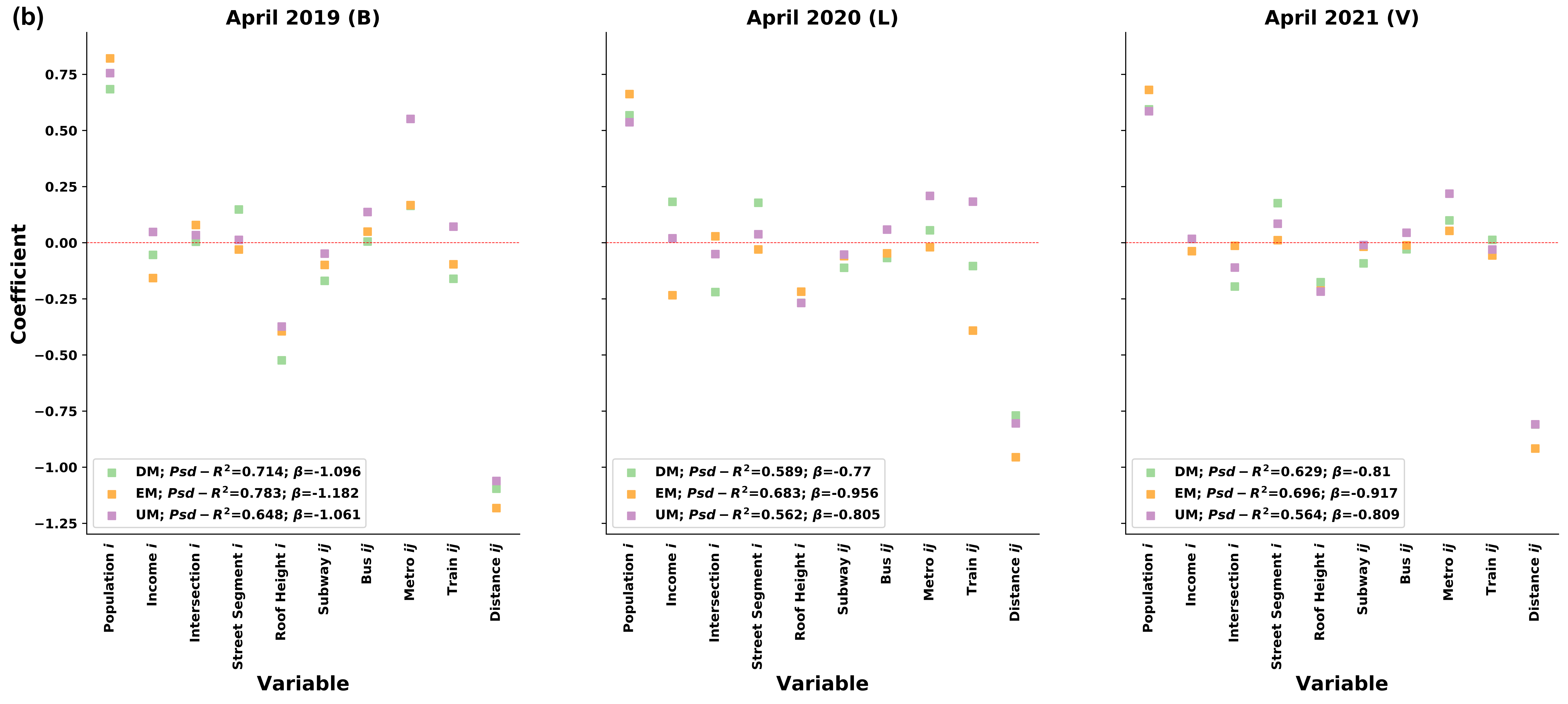}
        \caption{\textbf{Destination/Attraction Constrained Model} is initialised by fitting destination balancing factor $B_j$ to estimated flows to each destination in such a way that its total value equals to observed total flows at destination. Controlling for flows at destination strikingly uplift predictive power comparing to the unconstrained version in Fig.~\ref{fig:3}. However, adding more independent variables into Gravity Model (Fig.~\ref{fig:5}a) in order to build Urban System Model (Fig.~\ref{fig:5}b) contribute to subtle increase in pseudo $R^2$, only by 3.5\% at the highest in April 2019 for downward mobility (DM). All variables are statistically significant at $\alpha$ = 5\%.}
        \label{fig:5}       % Give a unique label
\end{figure*}

After constraining flows and other variables on destination side, the result of gravity model (Fig.~\ref{fig:5}a) and urban system model (Fig.~\ref{fig:5}b) is now comparable where  all variables are  statistically significant at $\alpha= 5\%$. It results in higher distance decay parameter $\beta$ than the unconstrained version and the origin constrained calibration. During the lockdown, distance effect is less deterrent due to restricted policy. The correction degree is more prevalent for downward mobility, as in the previous two model. Higher predictability level of equal mobility is also confirmed especially in the baseline period (pseudo $R^2 = 0.755$ for Gravity Model; pseudo $R^2 = 0.783$ for Urban System Model). In spite of that, augmenting more variables in urban system model under destination constrained setting only gives negligible prediction improvement in which the highest is retrieved in baseline period for downward mobility at 3.48\%. Accordingly, push factor attributes don't really uplift the flows.

\clearpage

\section{Discussion and conclusions}
\label{sec:disscussion}
\setlength{\parindent}{2em}
\setlength{\parskip}{1em}

This study aims to find the best fit for each specific setting in mobility and to quantify an extent external shock represented by COVID outbreak diverts the typical mobility structure. We propose combinatorial experimental design specifically target mobility stratification and later use this as a basis for estimating mobility flow. Given all possible pairings at 3 levels (Constraint-Period-Type) for 2 sets of variable in each model (Gravity Model and Urban System Model) as seen in Fig.~\ref{fig:3a} and Fig.~\ref{fig:3b}, evaluation of each estimation model realisation is performed by calculating pseudo $R^2$ as goodness of fit. This measure is selected due to its convenient interpretation that resembles $R^2$ in OLS and compatible specification for likelihood based function such as GLM \cite{oshan2016primer, faraway2016extending}. Within the range from 0 to 1, higher pseudo $R^2$ indicates better model fit. 

%\subsection{Model fit}
%\label{fit}
In summary, we propose urban system model as an extension to the gravity model. The results generated by urban system model then compared with the benchmark gravity model to. To capture prediction improvement contributed by the inclusion of additional variables, among others transportation network and spatial morphology, Fig.~\ref{fig:5}c summarises the difference in pseudo $R^2$ between gravity model Fig.~\ref{fig:5}a and urban system model Fig.~\ref{fig:5}b. Overall upper mobility attains 17.88\% improvement on average across constraints, on top of downward mobility (10.89\%) and equal mobility (8.15\%). Among other pairings, model extension works best for upper mobility under origin constrained treatment in vaccination period. The accuracy between predicted and observed flows as shown by pseudo $R^2$ is boosted by nearly 30\% after augmenting those variables into the model. 

\begin{figure*}[ht!]
    \centering
        \includegraphics[width=0.8\linewidth]{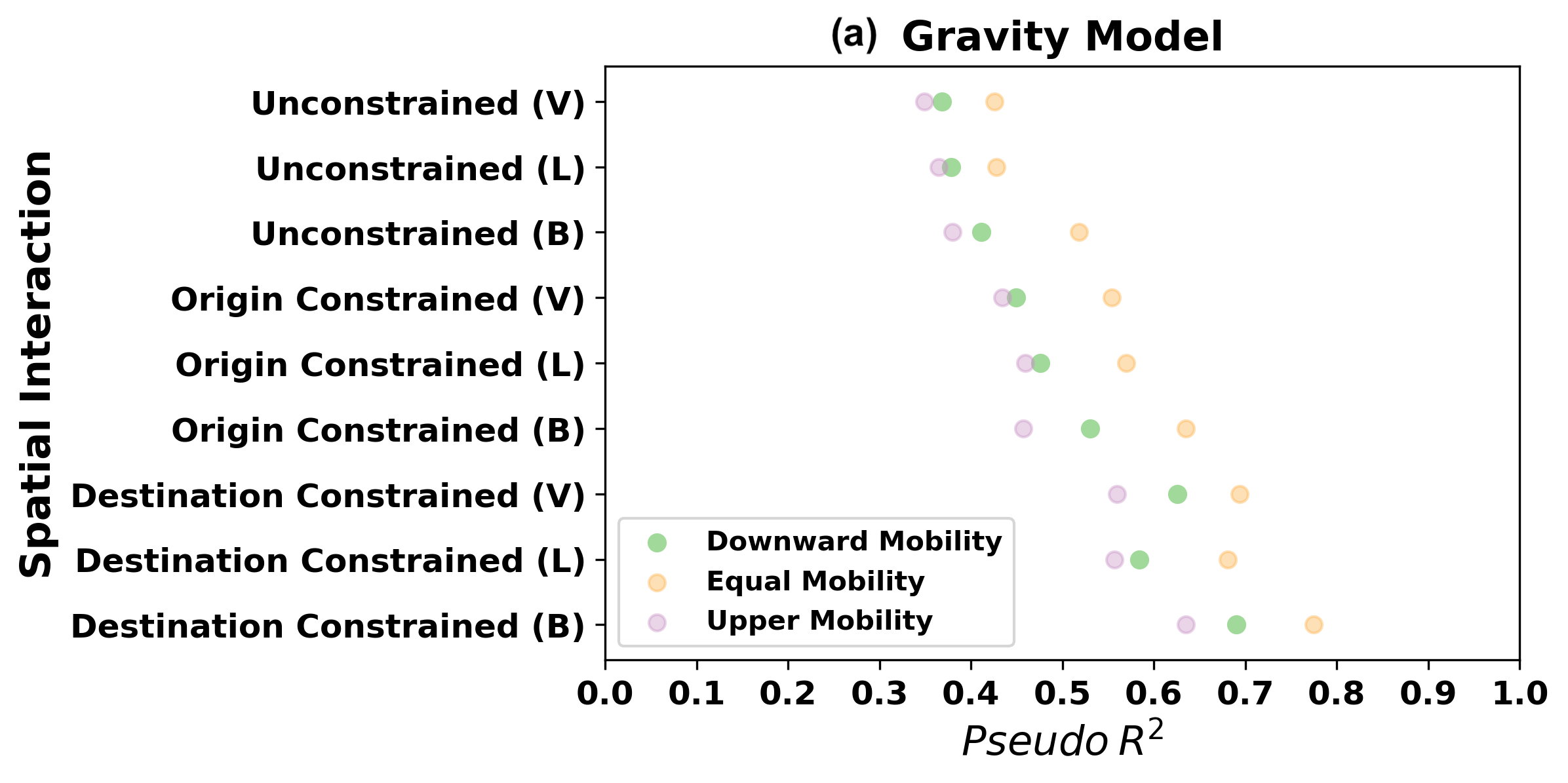}
        \includegraphics[width=0.8\linewidth]{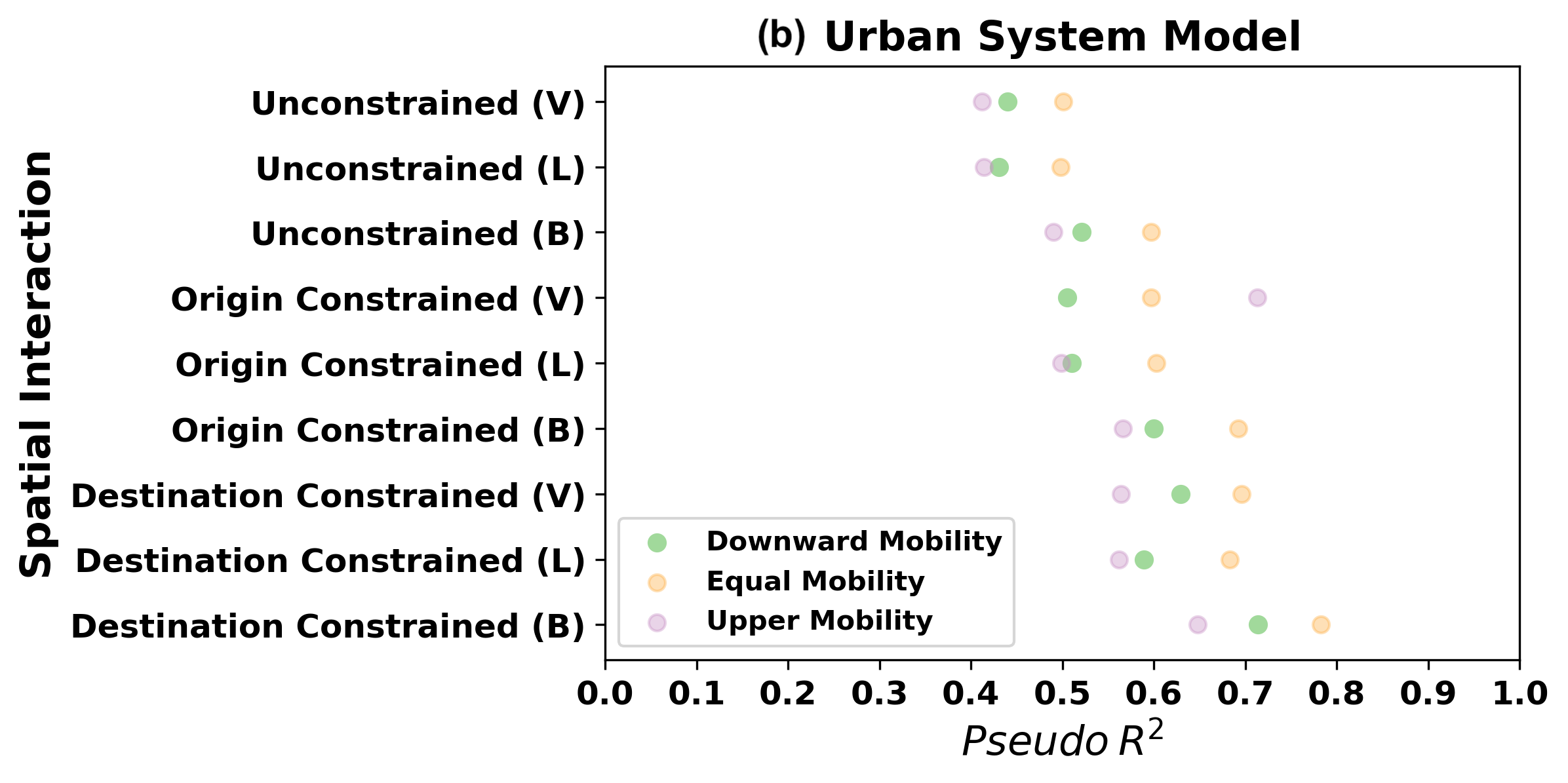}
        \includegraphics[width=0.8\linewidth]{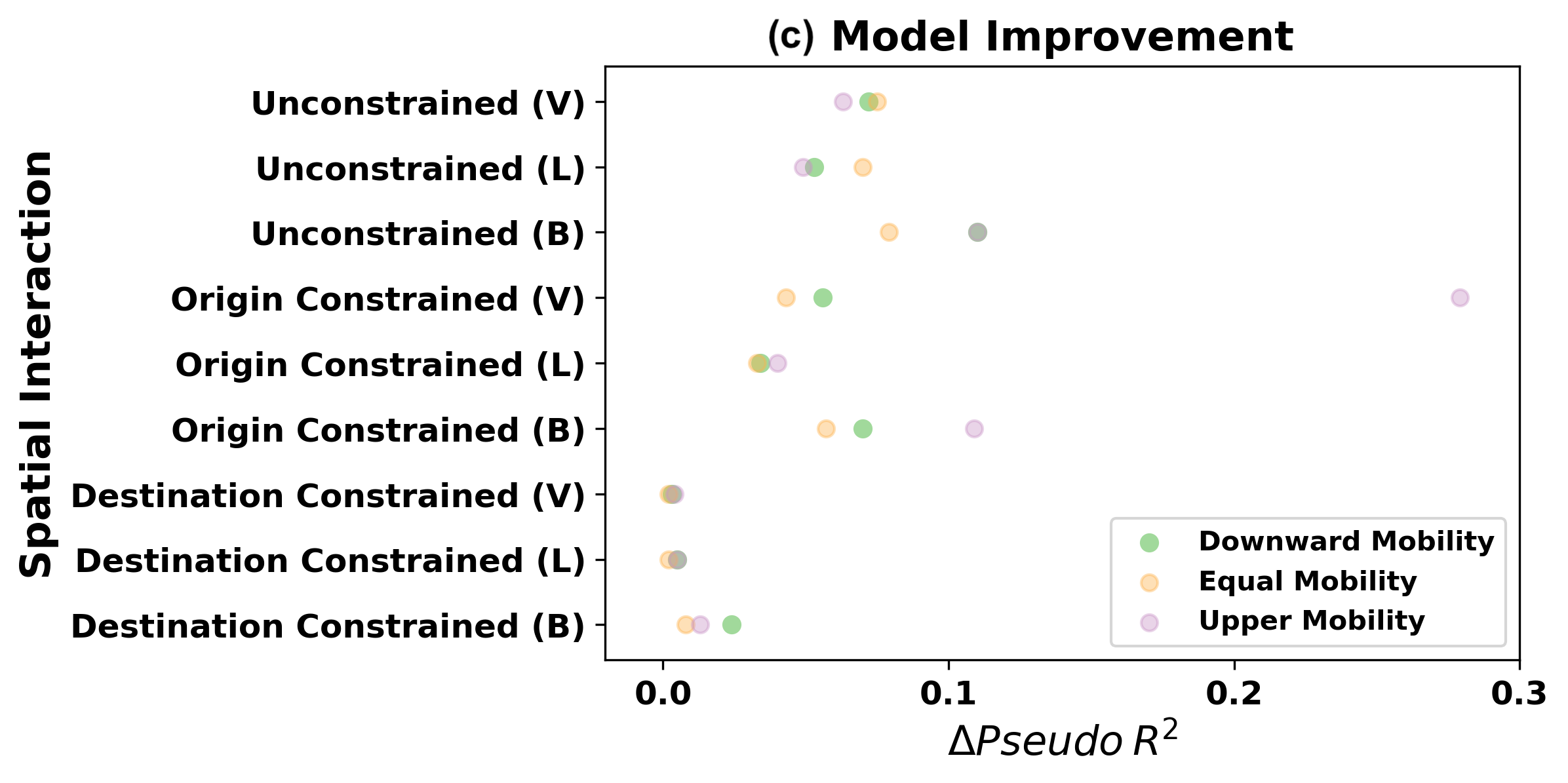}
        \caption{\textbf{Model fit (Pseudo $R^2$)} is presented for Gravity Model (Fig.~\ref{fig:6}a) and Urban System Model (Fig.~\ref{fig:6}b). Model improvement measured as the difference between the two (Fig.~\ref{fig:6}c) is at the highest for upper mobility under origin constrained treatment in vaccination period.}
        \label{fig:6}       % Give a unique label
\end{figure*}

Coherence between upward mobility and transportation network is rooted in two aspects. Firstly, the accessibility of public transportation network modulates commuting flows directing to areas with higher income, but not so much to areas within similar economic level. Mostly, such mobility is driven by commuting to workplaces as areas with better economic prospect offers more opportunities. Another plausible reason could be linked to the spatial distribution pattern as previously seen in Fig.~\ref{fig:1}c where SES distance is intertwined with geographic distance. Distance proximity in the case of income class similarity forms clustering tendency in which areas categorised in the same income class tends to be located next to each other at the given distance radius. To confirm the impact of transportation network in mobility stratification, upcoming study should test the magnitude of spatial autocorrelation distribution by applying Moran Moran's I index, for instance. 

%This finding is in conformity to the presence of push, pull, and interaction factors discussed further in Section \ref{factor}. 

%If we specifically look at the model parameters in this setting (purple blocks in Fig.~\ref{fig:4}b for April 2021),  

On the changing distance decay effect, urban system model  contributes to refining distance dimension. We take the average value of difference in $\beta$ between the two models across period for each mobility type. Distance becomes most deterrent in upper mobility under various constraint treatments: unconstrained (12.17\%), origin constrained (10.6\%), and destination constrained (3.74\%). It implies that controlling for origin attributes (Fig.~\ref{fig:4}b) and destination attributes(Fig.~\ref{fig:5}b) stabilise the the influence of distance in mobility as the two models give less obvious discrepancy on $\beta$. Less staggering distance decay effect is also found in downward and equal mobility.

We then focus on whether the prominence of pull and push effect are identical in the midst of period switch by looking at coefficient of each variables. In unconstrained spatial interaction, the strength of destination areas in attracting flows is more powerful than origin areas in triggering flows. This finding is also conclusively found in origin constrained spatial interaction, so does in destination constrained spatial interaction. Moreover, interaction factors represented by public transport connectivity between areas, more prominently metro, signifies the degree of mobility across temporal window.

\clearpage

\bibliographystyle{unsrt}
\bibliography{sample}

%\printbibliography

\end{document}